\documentclass[twoside,twocolumn,english,superscriptaddres,showpacs,longbibliography,aps,prl]{revtex4-2}

\usepackage[T1]{fontenc}
\usepackage[utf8]{inputenc}
\usepackage[english]{babel}
\usepackage{bm}
\usepackage{amsmath}
\usepackage{amssymb}
\usepackage{graphicx}
\usepackage{circuitikz}
\usepackage[unicode=true, pdfusetitle, bookmarks=true,bookmarksnumbered=true,bookmarksopen=true,bookmarksopenlevel=1,
breaklinks=false,pdfborder={0 0 0},pdfborderstyle={},backref=false,color links, citecolor=blue,linkcolor=red,urlcolor=blue] {hyperref}
\usepackage[nameinlink]{cleveref}
\Crefname{figure}{Fig.}{}

\makeatletter
\def\set@firstnote#1{%
 \@ifnum{\firstnote@num=#1\relax}{}{%
  \class@warn@end{Endnote numbers changed: rerun LaTeX}%
 }%
 \immediate\write\@mainaux{%
   \global\mathchardef\string\firstnote@num#1\relax
 }%
}%

\usepackage{braket}
\usepackage{bm}
\usepackage{tikz}
\usepackage{pgfplots}
\usepackage{pgf}
\usepackage{subfigure}
\usepackage{nicematrix}
\usepackage{diagbox}
\usepackage{orcidlink}

\renewcommand{\selectlanguage}[1]{}

\makeatother

\pgfplotsset{compat=1.18}
\begin{document}

\title{Classifying fusion rules of anyons or SymTFTs: A general algebraic formula for domain wall problems and quantum phase transitions}
\author{Yoshiki Fukusumi}
\affiliation{Physics Division, National Center for Theoretical Sciences, National Taiwan University, Taipei 106319, Taiwan}
\date{\today}
\begin{abstract}
We propose a formula for the transformation law of anyons in topologically ordered phases or topological quantum field theories (TQFTs) through a gapped or symmetry-preserving domain wall. Our formalism is based on the ring homomorphism between the $\mathbb{C}$-linear commutative fusion rings, also known as symmetry topological field theories (SymTFTs). The fundamental assumption in our formalism is the validity of the Verlinde formula, applicable to commutative fusion rings. By combining it with more specific data of the settings, our formula provides classifications of anyons compatible with developing categorical formulations. It also provides the massless renormalization group (RG) flows between conformal field theories (CFTs), or a series of measurement-induced quantum phase transitions, in the language of SymTFT, through the established correspondence between CFTs and TQFTs. Moreover, by studying the correspondence between the ideal structure in the massless RG and the module in the related massive RG, one can make the Nambu-Goldstone-type arguments for generalized symmetry. By combining our formula with orbifolding, extension, and similarity transformation, one can get a series of classifications for the corresponding extended models, or symmetry-enriched topological orders and quantum criticalities. 
\end{abstract}

\maketitle

\section{introduction}
\label{introduction}

Anyon\cite{Leinaas:1977fm,Goldin:1979ki,Goldin:1981sm,Wilczek:1981du} has a significant role in many aspects of contemporary physics (see also historical remarks in \cite{Goldin:2022gcn}), such as quantum phase transitions and topological orders\cite{Laughlin:1983fy,Wen:1989iv,Wen:1990se,moore_nonabelions_1991}, and quantum information\cite{Kitaev:1997wr,Kitaev:2006lla,nayak_non-abelian_2008}. 
Gapped or renormalization group (RG) domain wall\cite{Brunner:2007ur,Kitaev:2011dxc,Gaiotto:2012np,Lan:2014uaa,Hung:2015hfa,kong2015boundarybulkrelationtopologicalorders,Kong:2017hcw,Klos:2019axh,Zhao:2023wtg} is a fundamental theoretical framework producing classifications of anyons realized in conformal field theories (CFTs) or the corresponding topological quantum field theories(TQFTs)\cite{Witten:1988hf}. They are also expected to provide the condensation rule of anyons\cite{Bais:2008ni,Kong:2013aya} or hierarchical structure in topologically ordered states, typically in fractional quantum Hall states\cite{Halperin:1983zz,Halperin:1984fn,Jain:1989tx,Bernevig_2008,Bernevig2008PropertiesON,Hansson_2017} (see also a recent work \cite{Zhang:2024bye}). There exist several related but slightly different versions of definitions, but the most fundamental and common concepts are some canonical relationships between fusion rules of anyons by a morphism in abstract algebra or a (tensor) functor in category theory. In particular, based on the recent development in the corresponding symmetry topological field theory (SymTFT)\cite{Apruzzi:2021nmk,Fukusumi:2024ejk,Antinucci:2025fjp,Fukusumi:2025ljx}, the most reasonable formulation for the domain walls is the relationship through ring homomorphisms between fusion rules\cite{Neupert:2016pjk,Wan:2016php,Klos:2019axh,Shen:2019wop,Zhao:2023wtg,ng2025classificationmodulardatarank,Fukusumi:2025clr,Fukusumi:2025xrj}. This relation itself is also a natural structure to obtain the functor in $\mathbb{C}$-linear category established in \cite{Petkova:2000ip}, where $\mathbb{C}$ is the field of complex numbers (For readers interested in related historical aspects, we note \cite{Evans:2023nbp}). There exists a reason to formulate the domain walls between $\mathbb{C}$-linear theories with a connection to the fundamental problem in realizing fractional quantum Hall systems in experimental settings\cite{Wang2017TopologicalOF,Mross_2018}. The results in \cite{Wang2017TopologicalOF,Mross_2018} indicate that the bulk-edge correspondence \cite{Laughlin:1983fy,Witten:1988hf,Moore:1991ks} predicted from the thermal Hall conductance\cite{Kane:1996bje} should be defined up to some connectivity of models\cite{Stone:2012ud,Son:2015xqa,Barkeshli:2015afa}. Consequently, the experimental realizations indicate the changes of chiral central charges between underlying CFTs  realized in the edge modes of topological orders, and this change of central charge is prohibited without the $\mathbb{C}$-linear formulations\cite{Moller:2024plb,Moller:2024xtt}. Regardless of their experimental importance, the $\mathbb{C}$-linear formulations of the domain walls between CFTs or TQFTs are relatively less familiar, and still in development\cite{Zhao:2023wtg,Fukusumi:2025clr,Fukusumi:2025xrj,Fukusumi:2025fir}. Moreover, corresponding to the complex number coefficients before anyonic objects, the simplicity or locality of objects can be changed in $\mathbb{C}$ linear theories. This is consistent with the anomaly inflow mechanism\cite{Callan:1984sa}, which permits the existence of exotic particles in the domain wall phenomena.

Hence, we restrict our attention to the $\mathbb{C}$ linear formalism with emphasis on \emph{linear algebra}, providing a straighforward quantum mechanical description, whereas there exist other interesting formalisms respecting bulk and defect RG flows. For example, we note that the other formalism has a significant benefit in studying the defect or boundary critical phenomena\cite{Dorey:2004xk,Dorey:2005ak,Green:2007wr,Fredenhagen:2009tn,Dorey:2009vg,Chang:2018iay,Ambrosino:2025pjj} with a connection to the anomaly inflow mechnism\cite{Callan:1984sa}. However, their precise relation to the (tensor) functor in category theory is less clear at this stage \footnote{We thank Gerard Watts and Ingo Runkel clarifying this point.}.

\begin{figure}[htbp]
\begin{center}
\includegraphics[width=0.5\textwidth]{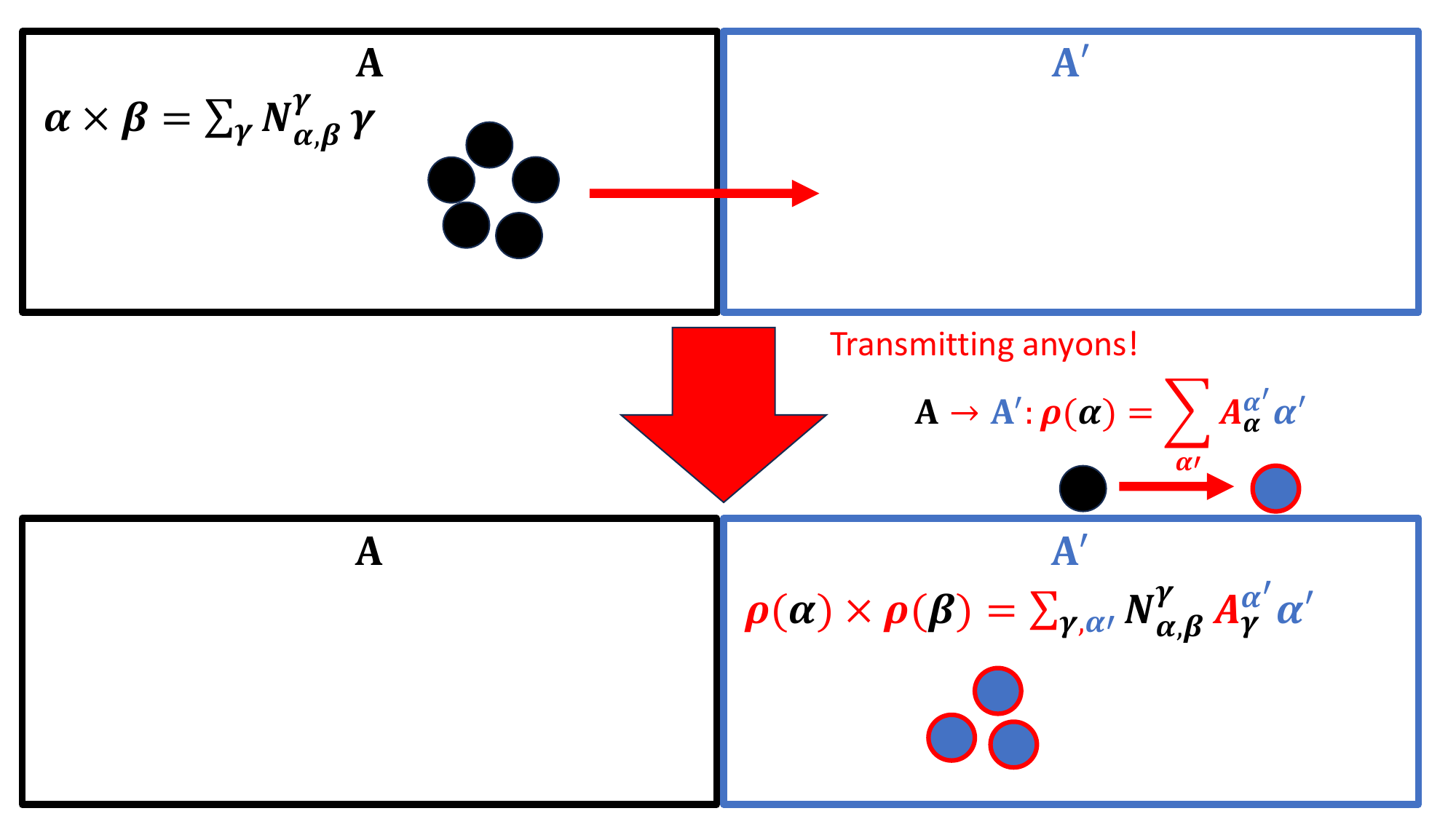}
\caption{ Picture of the transformation of anyons through a domain wall. The UV objects and structures are represented in black, and those of the IR are represented in blue. The red colour represents the homomorphism or domain wall transforming anyons. Because of the surjective property of the homomorphism, the number of types of anyon decreases under the application of the homomorphism. The fundamental point is that the fusion and linear sum, $\times$ and $+$ (or $\sum$), are compatible with the application $\rho$. One can see related figures in \cite{Fukusumi:2025clr,Fukusumi:2025xrj}, but we do not assume the coset or level-rank duality structures in the present paper.
}
\label{domain_wall_transformation}
\end{center}
\end{figure}

In most of the references studying the gapped domain walls, either $\mathbb{C}$-linear or not, the properties of ring homomorphism $\rho: \mathbf{A}\rightarrow \mathbf{A'}$ are \emph{assumed} and their consequences have been studied where $\mathbf{A}=\{ \alpha\}_{\alpha}$ and $\mathbf{A'}=\{ \alpha'\}_{\alpha'}$ are the corresponding ultraviolet (UV) and infrared (IR) fusion ring and $\alpha$ and $\alpha'$ are their anyons or symmetry operators. We also note that we do not assume the chirality or orientation of $\mathbf{A}$ and $\mathbf{A'}$ (see \cite{Kong:2019cuu,Fukusumi:2024ejk} for related discussions). The homomorphism provides the transformation law of anyon $\rho(\alpha)=\sum_{\alpha'}A_{\alpha}^{\alpha'}\alpha'$ (Fig. \ref{domain_wall_transformation}). However, one can notice a puzzle in this research direction: How can one construct $\rho$ or determine the coefficients $A_{\alpha}^{\alpha'}$? We remind that the ring homomorphism is a linear mapping compatible with $\times$ operations, i.e. $\rho(\alpha\times \beta)=\rho(\alpha)\times \rho (\beta)$, and it has been clarified that the exact algebraic construction of $\rho$ requires complicated calculations even in familiar models\cite{Fukusumi:2025clr,Fukusumi:2025xrj}.  If the UV and IR theories are related by the coset construction\cite{Goddard:1984vk,Goddard:1984hg,Goddard:1986ee} or level-rank duality\cite{Kuniba:1990im,Nakanishi:1990hj,Aharony:2016jvv}, one can solve this problem straightforwardly by mapping the problem to a domain wall problem in TQFTs and projecting out the particle at the domain wall\cite{Fukusumi:2025xrj}. Related to this, in several research communities, it is known that some correspondence between the massless renormalization group (RG) flows\cite{Zamolodchikov:1987ti,Zamolodchikov:1987jf,Zamolodchikov:1989hfa} and the coset construction in CFTs should hold \cite{LeClair:2001yp,Lecheminant:2015iga,Sfetsos:2017sep,Georgiou:2018gpe,Borsato:2023dis,Cordova:2025eim}. One can see the related discussions for lattice-gauge models in \cite{Zhao:2025zsb}. However, in general, the construction of a homomorphism is a difficult problem and requires extensive categorical or combinatorial calculations\cite{Cong:2016ayp,Cong:2017ffh,Chen:2025xbp,Jia:2025yph,Li:2025omz}. We also note related studies on conformal interfaces as the analytical analog \cite{Quella:2006de,Gaiotto:2012np,Kimura:2014hva,Kimura:2015nka,Stanishkov:2016pvi,Stanishkov:2016rgv,Poghosyan:2022mfw,Poghosyan:2022ecv,Poghosyan:2023brb,Cogburn:2023xzw,Fukusumi:2024ejk,Antinucci:2025uvj,Furuta:2025ahl,Fukusumi:2025ljx}. Historically, this research direction has been explored in the Kondo or Kane-Fisher problem (or related wire-junction problem) in condensed matter\cite{Kondo:1964nea,Affleck:1990by,Kane:1992xse,Kane:1992zza,Wong:1994np,Affleck:1995ge,Saleur:1998hq}.

In this paper, we propose a general formula, Eq.\eqref{Verlinde-like_formula}, for constructing homomorphisms between a wider class of fusion rings. Technically, our method requires some elementary calculations in abstract algebra (or linear algebra)\footnote{This fundamental fact in abstract algebra has been notified to the author by Gerard Watts and Ingo Runkel.}. Hence, we only note a classic textbook\cite{atiyah1969introduction}. This simplicity is in sharp contrast with other frameworks. The most fundamental assumption in this manuscript is the \emph{Verlinde formula}\cite{Verlinde:1988sn} (in commutative fusion rings) and the underlying modular $S$ properties of the theories. The resultant formula for the homomorphism is also similar to the Verlinde formula, and it provides a possible form of homomorphism between fusion rings. For readers interested in mathematical or categorical formulation, we note that the formula provides algebraic data of a functor between $\mathbb{C}$-linear categories.

The rest of the manuscript is organized as follows. In Sec. \ref{classification}, we introduce the general formula, Eq.\eqref{Verlinde-like_formula}, constructing ring homomorphisms between fusion rings. This is the main result of this paper, and the formula applies to any categorical framework if the corresponding functor is well-defined. In Sec. \ref{unbroken}, we discuss the more involved classification by conformal spin of anyons or associated sectors after constructing the homomorphism. The implications for the anomaly cancellation mechanism and related categorical notions are also discussed. We study the correspondence between the algebraic structure, \emph{ideal} in mathematics, in massless RG and module in the massive RG and its analogy to Nambu-Goldstone type phenomenology in Sec. \ref{Nambu-Goldstone}. Sec. \ref{conclusion} is the concluding remark, and we summarize the consequent phenomenology and comment on the future research directions. In End Matter, we discuss the application of our method to more general settings following research directions in \cite{Fukusumi:2024ejk,Fukusumi:2025ljx,Fukusumi:2025fir}. The results are summarized in Eq.\eqref{sequence} and Eq.\eqref{RG_similarity}.

\section{Classification of anyons or SymTFTs}
\label{classification}

In this section, we provide a complete classification of commutative rings and their representation by SymTFT under several assumptions. Eq.\eqref{Verlinde-like_formula} is the main result in this section. We note a few reviews and textbooks\cite{Ginsparg:1988ui,Fuchs:1992nq,DiFrancesco:1997nk,Ribault:2016sla} for general references in two-dimensional CFTs and several reviews\cite{Cordova:2022ruw,McGreevy:2022oyu,Schafer-Nameki:2023jdn,Shao:2023gho} for references on generalized symmetry \cite{Cobanera:2009as,Cobanera:2011wn,Cobanera:2012dc,Gaiotto:2014kfa}(See also a concise essay \cite{Chen:2025uno}). First, we assume the following data:
\begin{itemize}
\item{There exist $n$ and $n’$ types of simple objects in the UV and IR theories, respectively.}
\item{The theories satisfy the Verlinde formula determined by the modular $S$ matrices.}
\item{The objects correspond to topological symmetry operators in 2d CFTs or anyons in 3d TQFTs, and form a $\mathbb{C}$-linear fusion ring as in \cite{Petkova:2000ip}.}
\end{itemize}
In the following discussion, we distinguish the UV and IR theories by specifying the prime symbol ’ to the IR theory. We denote the algebraic objects in SymTFTs in CFTs or anyons in TQFTs by the Greek symbols $\alpha, \beta,…$. For simplicity, we assume that the theories are diagonal $A$-type model with the following modular invariant,
\begin{equation}
Z(\tau)=\sum_{\alpha}|\chi_{\alpha}|^{2}
\end{equation}
where $\chi$ is the chiral character and $\tau$ is the modular parameter. The matrix representation of the modular $S$ transformation is $\chi_{\alpha}(-1/\tau)=\sum_{\beta}\chi_{\beta}S_{\alpha,\beta}$ and this plays the fundamental role in studying the fusion rings via the Verlinde formula, as we discuss below.

Then, we introduce the following fusion ring,
\begin{equation}
\alpha \times \beta=\sum_{\gamma}N_{\alpha,\beta}^{\gamma}\gamma,
\end{equation}
where $N$ is determined by modular $S$ matrix by the Verlinde formula \cite{Verlinde:1988sn},
\begin{equation}
N^{\gamma}_{\alpha,\beta}=\sum_{\delta}\frac{S_{\alpha,\delta}S_{\beta,\delta}\overline{S_{\delta,\gamma}}}{S_{I,\delta}},
\label{Verlinde_formula}
\end{equation}
where $I$ is the identity operator corresponding to the vacuum and $N$ is the fusion matrix $\alpha \times \beta =\sum_{\gamma} N^{\gamma}_{\alpha,\beta}\gamma$. In this setting, one can represent the anyons by idempotents, ${ e{(\alpha)} }$\cite{Petkova:2000ip},
\begin{equation}
\alpha=\sum_{\beta} \frac{S_{\alpha,\beta}}{S_{I, \beta}}e_{(\beta)}
\end{equation}
such as $e_{(\alpha)}\times e_{(\beta)}=\delta_{\alpha,\beta}e_{(\alpha)}$. Inversely, one can represent the idempotent from the anyon as,
\begin{equation}
e_{(\alpha)}=\sum_{\beta}S_{I,\alpha}\overline{S_{\alpha,\beta}} \beta.
\end{equation}
For the Verlinde lines, one can introduce the corresponding idempotents as projections\cite{Petkova:2000ip};
\begin{equation}
e_{(\alpha)}=P_{\alpha}=\sum_{M, \overline{M}}\left(|\alpha,M\rangle \langle \alpha, M|\right)\otimes\left(\overline{|\alpha,\overline{M}\rangle} \overline{\langle \alpha, \overline{M}|}\right)
\end{equation}
where $M$ ($\overline{M}$) is the label of (anti)chiral descendant states, producing the (anti)chiral character $\chi_{\alpha}$ ($\overline{\chi_{\alpha}})$ by taking their traces. More generally, one can introduce similar chiral or antichiral symmetry operators\cite{Chang:2018iay,Komargodski:2020mxz,Lin:2022dhv,Fukusumi:2025ljx}, and we discuss related involved problems resulting from the chiralities or braiding properties in Sec. \ref{unbroken}. However, the algebraic aspects are robust against such subtleties.

For the latter discussion, we note the benefits of introducing the idempotents. The idempotent representation provides algebraic data of a ring homomorphism straightforwardly. It should be noted that the anyon basis provides the natural phenomenology or intuitions, but their classifications by homomorphism are difficult to read. This may be a reason why the explicit coefficient of the gapped domain wall has not been calculated in the literature, and the resolution of this difficulty is the main result of the present paper. We also note that the Verlinde formula and related categorical notions appeared with a close connection to the boundary or defect CFT\cite{Cardy:1989ir,Behrend:1999bn} (see also reviews\cite{Petkova:2000dv,Cardy:2004hm,Schweigert:2000ix} and a textbook\cite{Recknagel:2013uja}). We note earlier references on the related subfactor theory\cite{article,Bockenhauer:1999wt} and related recent reviews\cite{Kawahigashi:2021hds,Evans:2023nbp}. However, we stress again that the $\mathbb{C}$-linear properties, the representation of fusion algebra as conserved charges, has been clarified in \cite{Petkova:2000ip}.

By using the idempotents, one can directly obtain homomorphisms by specifying the preserved idempotents $\{ e_{(\alpha_{\text{pr}})}\}$, as follows:
\begin{itemize}
\item{The preserved idempotent $\mathbb{S}_{\rho}$ is mapped to some idempotent in the IR theory, $\rho(e_{(\alpha_{\text{pr}})})=e_{\alpha'_{\text{pr}}}$.}
\item{The unpreserved idempotent $e_{(\alpha)}$, $(\alpha)\notin \mathbb{S}_{\rho}$, is mapped to $0$ in the IR theory, $\rho(e_{(\alpha)})=0$.}
\end{itemize}
Because of the linearlity, $\rho(\alpha+\beta)=\rho(\alpha)+\rho(\beta)$, and the compativility of the product, $\rho(e_{(\alpha)}\times \rho_{(\beta)})=\rho(e_{(\alpha)})\times \rho(e_{(\beta)})$, the above mapping provides the ring homomorphism, compativility of $+$ and $\times$ operations under the application $\rho$. 
For latter use, we note the set of the label of preserved idempotent as $\{ (\alpha_{\text{pr}})\}=\mathbb{S}_{\rho}$ and the set of the UV and IR pairs as $\mathbb{S}_{\rho,\times}=\{ (\alpha_{\text{pr}}, \alpha'_{\text{pr}})\}$. We represents $\mathbb{S}_{\rho}$ as linear subspace, pseudo-ring, or set of the labels depending on the contexts, and we denote the broken (or unpreserved) sectors as $\mathbb{S}_{\rho}^{\text{c}}=\{ e_{(\alpha)}\}_{\alpha\notin \mathbb{S}_{\rho}}$.

By specifing the preseved idempotents $\mathbb{S}_{\rho, \times}=\{ (\alpha_{\text{pr}},\alpha'_{\text{pr}})\}$, one can summarize the homomorphism as 
\begin{equation}
\begin{split}
\rho(e_{(\alpha_{\text{pr}})})&=e_{(\alpha'_{\text{pr}})} \\
\rho(e_{(\alpha)})&=0, \text{for $\alpha \notin \mathbb{S}_{\rho}$} 
\end{split}
\end{equation}
Hence, if the UV and IR theories have $n$ and $n'$ objects respectively, one will obtain a $P(n,n')$ set of surjective homomorphisms. For example, when studying the flow from $M(2,7)$ minimal to $M(2,5)$ minimal model, there exist $6=3\times 2$ homomorphisms, and this number $6$ matches with the result in \cite{Fukusumi:2025xrj}. In the following, we apply the linear transformation from the idempotent basis and anyon basis for further  understanding. 

First, from the compatibility between the linear sum and homomorphism, the following holds,
\begin{equation}
\rho(\alpha)=\sum_{\beta} \frac{S_{\alpha,\beta}}{S_{I, \beta}}\rho\left(e_{(\beta)} \right)
\end{equation}
Next, we apply the homomorphism to each idempotent and obtain the form,
\begin{equation}
\begin{split}
\sum_{\beta} \frac{S_{\alpha,\beta}}{S_{I, \beta}}\rho\left(e_{(\beta)} \right) 
&=\sum_{\beta_{\text{pr}}\in \mathbb{S}_{\rho}}\frac{S_{\alpha,\beta_{\text{pr}}}}{S_{I, \beta_{\text{pr}}}}\rho\left(e_{(\beta_{\text{pr}})} \right) \\
&=\sum_{(\beta_{\text{pr}}, \beta'_{\text{pr}})\in \mathbb{S}_{\rho,\times}}\frac{S_{\alpha,\beta_{\text{pr}}}}{S_{I, \beta_{\text{pr}}}}e_{(\beta'_{\text{pr}})}, 
\end{split}
\end{equation}
By applying the basis transformation from idempotents to anyons in the IR, we obtain the  form,
\begin{equation}
\begin{split}
&\sum_{(\beta_{\text{pr}}, \beta'_{\text{pr}})\in \mathbb{S}_{\rho,\times}}\frac{S_{\alpha,\beta_{\text{pr}}}}{S_{I, \beta_{\text{pr}}}}e_{(\beta'_{\text{pr}})} \\
&=\sum_{(\beta_{\text{pr}}, \beta'_{\text{pr}})\in \mathbb{S}_{\rho,\times}}\frac{S_{\alpha,\beta_{\text{pr}}}}{S_{I, \beta_{\text{pr}}}}\left(\sum_{\alpha'}S_{I',\beta'_{\text{pr}}}\overline{S_{\beta'_{\text{pr}},\alpha'}} \alpha'\right)
\end{split}
\end{equation}
Consequently, we obtain the following formula for the general homomorphism,
\begin{equation}
\rho(\alpha)=\sum_{\alpha', (\beta_{\text{pr}}, \beta'_{\text{pr}})\in \mathbb{S}_{\rho,\times}}\frac{S_{\alpha,\beta_{\text{pr}}}S_{I',\beta'_{\text{pr}}}\overline{S_{\beta'_{\text{pr}},\alpha'}} }{S_{I, \beta_{\text{pr}}}}\alpha'.
\label{homomorphism_formula}
\end{equation}
The coefficients of the homomorphism or splitting property of anyons, $\rho(\alpha)=\sum_{\alpha'}A_{\alpha}^{\alpha'} \alpha'$, can be represented as,
\begin{equation} 
A_{\alpha}^{\alpha'}=\sum_{(\beta_{\text{pr}}, \beta'_{\text{pr}})\in \mathbb{S}_{\rho,\times}}\frac{S_{\alpha,\beta_{\text{pr}}}S_{I',\beta'_{\text{pr}}}\overline{S_{\beta'_{\text{pr}},\alpha'}} }{S_{I, \beta_{\text{pr}}}}
\label{Verlinde-like_formula}
\end{equation}
This form itself is analogous to the Verlinde formula but it contains both contributions of the UV and IR structures. One can see related discussions in \cite{Shen:2019wop,Zhao:2023wtg,ng2025classificationmodulardatarank}, but our method only requires a modular $S$ matrix of the UV and IR theories and is more general and concise. More or less, the existing methods usually require additional constraints such as the properties of domain walls, and these constraints restrict possible domain walls very strongly. For example, it is impossible to connects two theories with different central charges in the original arguments on the gapped domain wall\cite{Kitaev:2011dxc,Lan:2014uaa}. Hence this formulation is insufficient to explain the controvercies of thermal Hall conductance\cite{Wang2017TopologicalOF,Mross_2018}, experimental realization of anomaly inflow mechanism\cite{Callan:1984sa,Stone:2012ud}. In \cite{Zhao:2023wtg}, the domain wall has been extended to a $\mathbb{C}$-linear functor, but still additional constraints from the so-called domain wall $S$-matrix appear. However, except for \cite{Zhao:2023wtg}, the irrational coefficients before objects are usually not permitted  (see \cite{Neupert:2016pjk,Shen:2019wop,ng2025classificationmodulardatarank}, for example), and this prohibits one from studying existing simple RG flows, such as the RG flow from the tricritical Ising model to the Ising model, in their formalisms\cite{Fukusumi:2025clr}. Hence, we propose Eq. \eqref{Verlinde-like_formula} which requires the weaker assumptions, to test the validity of the ring homomorphism and related ideas in future studies.

In our formalism, one can easily detect the structure of an \emph{ideal} as idempotent $\mathbb{S}_{\rho}^{\text{c}}=\{ e_{(\alpha)}\}_{\alpha\notin \mathbb{S}_{\rho}}$. An ideal $\mathbf{I} (\subset \mathbf{A})$ is a linear subspace of $\mathbf{A}$ such as $\mathbf{A}\times \mathbf{I}=\mathbf{I}$, and $\mathbb{S}_{\rho}^{\text{c}}$ satisfy this condition (we note a classic paper\cite{Noether1921} and its English translation\cite{berlyne2014idealtheoryringstranslation}). For more detailed explanations, let us assume the perturbative realization of the massless flow $\rho$ as $H_{\text{UV}}+\lambda H_{\text{pert},\rho}$ where $H_{\text{UV}}$ is the Hamiltonian of the UV theoery, $H_{\text{pert}, \rho}$ is the perturbation triggering the flow $\rho$, and $\lambda$ is the coupling constant. When identifying the massless flow as $\lambda>0$,  there will exist the corresponding dual massive RG with $\lambda<0$. More symbolically, the decomposition of the UV theory, $\mathbf{A}=(\mathbf{A}/\mathbb{S}_{\rho}^{\text{c}})\oplus \mathbb{S}_{\rho}^{\text{c}}$ provides the decomposition of the low-energy sector as the quotient ring $\mathbf{A}/\mathbb{S}_{\rho}^{\text{c}}$ and the high-energy sector as ideal $\mathbb{S}_{\rho}^{\text{c}}$ in the massless flow $\rho$. By applying the Noether's ring isomorphism theorem, one can identify $\mathbf{A'}$ as $\mathbf{A}/\mathbb{S}_{\rho}^{\text{c}}$, and the low energy sectors correctly reproduce the consequence of $\rho$. Under the dual operation $\lambda \Rightarrow - \lambda$, the roles of high and low energy sector are exchaged.

Moreover, it is known both combinatorially and phenomenologically that the ground states of the massive RG can be described by the smeared boundary states\cite{Date:1987zz,Saleur:1988zx,Cardy:2017ufe,Lencses:2018paa}. Hence, by identifying the ideal as module appearing in the massive RG as in \cite{Cardy:2017ufe,Thorngren:2019iar,Fukusumi:2025clr,Fukusumi:2025xrj}, one can obtain the module of massive RG which is dual to the massless flow as $\{ |\alpha\rangle \rangle\}_{\alpha\notin \mathbb{S}_{\rho}}$, where $| \alpha \rangle \rangle$ is the (smeard) Ishibashi state\cite{Ishibashi:1988kg}, satisfying the relation $\langle\langle \alpha| e^{i\pi\tau H_{CFT}}|\beta \rangle \rangle=\chi_{\alpha}(\tau)\delta_{\alpha,\beta}$ where $H_{CFT}$ is the CFT Hamiltonian, $\tau$ is the modular parameter and $\delta_{\alpha,\beta}$ is the Kronekker delta. We also note that this mapping from $\mathbb{S}_{\rho}^{\text{c}}$ to $\{ |\alpha\rangle \rangle\}_{\alpha\notin \mathbb{S}_{\rho}}=\{ |\alpha\rangle \rangle\}_{\alpha\in \mathbb{S}_{\rho}^{\text{c}}}$ is called \emph{fiber} functor, and the symbol $\times$ is \emph{forgotten} under the mapping. In category theory, this is called a forgetful functor\footnote{However, one can \emph{remember} $\times$from Cardy's conditions}. By applying linear transformations, one can obtain the corresponding representation of (smeared) Cardy states\cite{Cardy:2017ufe} as 
\begin{equation}
\{ |\alpha\rangle \rangle\}_{\alpha\notin \mathbb{S}_{\rho}}=\{ \sqrt{S_{I, \alpha}}\sum_{\beta} \overline{S_{\alpha,\beta}}|\beta \rangle\}_{\alpha\notin \mathbb{S}_{\rho}} 
\end{equation}
where $|\alpha\rangle$ is the Cardy states determined by the linear relation, 
$|\alpha\rangle=\sum_{\beta}\frac{S_{\alpha,\beta}}{\sqrt{S_{I,\beta}}}|\beta\rangle\rangle$ and satisying the Cardy's condition for the annulus partition function,  $Z_{\alpha,\beta}(\tau)=\langle \alpha| e^{i\pi\tau H_{CFT}}|\beta \rangle =\sum_{\gamma}N_{\alpha,\beta}^{\gamma}\chi_{\gamma}(-1/\tau)$. We note several reviews and textbooks\cite{Cardy:2004hm,Petkova:2000dv,Recknagel:2013uja,Northe:2024tnm} for more detailed explanations of BCFTs, and the pioneering work \cite{Graham:2003nc} by Graham and Watts for more detailed symmetry analysis on the resultant modules. The fundamental algebraic relation derived in \cite{Graham:2003nc} is,
\begin{equation}
\mathcal{Q}_{\alpha} |\beta\rangle=\sum_{\gamma}N^{\gamma}_{\alpha,\beta} |\gamma\rangle
\end{equation}
where $ \mathcal{Q}_{\alpha} = \sum_{\beta} \frac{S_{\alpha,\beta}}{S_{I, \beta}}P_{\beta}$ is the topological symmetry operator (or conserved charge) satisfying the same fusion rule of $\{ \alpha \}$. One can see related discussion in \cite{Thorngren:2019iar,Kikuchi:2022gfi,Kikuchi:2023gpj,Fukusumi:2024ejk,Fukusumi:2025clr,Wen:2025xka,Choi:2025ebk,Zhang:2025xkw,Fukusumi:2025xrj}. For readers interested in the historical aspect of the correspondence between BCFT and massive RG, we note earlier works in massive integrable models \cite{Date:1987zz,Saleur:1988zx} (See also the historical remarks in \cite{Foda:2017vog,Fukusumi:2024ejk,Choi:2025ebk}). In condensed matter theory, the resultant correspondence between BCFT (or chiral CFT) and the bulk states of the topological orders can be seen in the Li-Haldane conjecture\cite{Li_2008} as has been clarified in \cite{Qi_2012}. More traditionally, this can be regarded as the cut and gluing properties of TQFTs and underlying CFTs with a close analogy to the tunneling problem\cite{Wen:1990se}.

In the subsequent discussions, we propose a more intuitive (or physical) understanding of the homomorphism and possible generalizations combined with related to other works by the author\cite{Fukusumi:2024cnl,Fukusumi:2024ejk,Fukusumi:2025ljx,Fukusumi:2025fir}.

\section{Preserved sectors and unbroken subalgebra}
\label{unbroken}
Whereas we have constructed the general class of ring homomorphisms, their properties (or physical implications) will depend on more detailed data or the setting of the theory, such as central charge and conformal spin of the objects. More precisely, the anomaly-free conditions of the unbroken subalgebra, $\mathbf{A}_{\text{ub}} \subset \mathbf{A}$ with $\mathbf{A}_{\text{ub}}\cap \text{Ker}\rho =\{ 0\}$, play fundamental roles in studying lattice or quantum field theoretic realizations. The condition $\mathbf{A}_{\text{ub}}\cap \text{Ker}\rho =\{ 0\}$ implements the ring isomorphism $\mathbf{A'}_{\text{ub}}=\rho\left(\mathbf{A}_{\text{ub}}\right)= \mathbf{A}_{\text{ub}}$, and this defines the preserved or unbroken symmetry structures in the homomorphism $\rho$. The folding trick\cite{Wong:1994np} is a standard method to map the problem of the homomorphism $\rho: \mathbf{A}\rightarrow \mathbf{A'}$ to the related BCFT of the tensor product of the coupled model $\mathbf{A} \otimes \overline{\mathbf{A'}}$. When the $\mathbb{Z}_{2}$ group structure, $\{ I, \psi\}$ at UV and $\{ I', \psi'\}$ at IR, is preserved, i. e. $\mathbf{A}_{\text{ub}}=\mathbb{Z}_{2}$, the extension of the $\mathbf{A} \otimes \overline{\mathbf{A'}}$ by the $\mathbb{Z}_{2}$ group ring $\{ I\otimes \overline{I'}, \psi \otimes \overline{\psi'}\}(\subset \mathbf{A} \otimes \overline{\mathbf{A'}})$ naturally encodes the data of RG flows\cite{Crnkovic:1989ug,Gaiotto:2012np}. More generally, for unbroken group ring structures, related discussion can hold \cite{Kong:2019cuu,Fukusumi:2024ejk}. Through the CFT/TQFT correspondence, this is nothing but a generalization of Lieb-Shultz-Mattis type arguments\cite{Lieb:1961fr,Schultz:1964fv} for gapless phases\cite{Furuya:2015coa,Lecheminant:2015iga,Yao:2018kel,Tanizaki:2018xto}. One can see related discussions in \cite{Antinucci:2025fjp}. 

More generally, when $\mathbf{A}_{\text{ub}}$ coincides with the connected etale algebra (or the condensable algebra\cite{Kong:2013aya,Zhang:2024bye}) of $\mathbf{A} \boxtimes\overline{\mathbf{A'}}$ where $\boxtimes$ is the Deligne product, the anomaly-free condition will be satisfied\cite{Kong:2019cuu,Kaidi:2021gbs,Fukusumi:2024ejk}. Because the connected etale algebra has braiding data outside of ring theoretic data, the classification (or the naturality of theories) can be more involved. For example, even when $\mathbf{A}_{\text{ub}}({\subset} \mathbf{A})$ and $\mathbf{A'}_{\text{ub}}(=\rho\left(\mathbf{A}_{\text{ub}}\right)\subset \mathbf{A'})$ are identical at the ring theoretic level, but after introducing the braiding properties or constraint from the conformal spin, the classifications can be different. It should also be noted that the systematic classification of the connected etale algebra of the chiral-chiral or chiral-antichiral product of CFTs is still in development. We note recent works on the unbroken subalgebraic structure in massless RGs\cite{Nakayama:2024msv,Kikuchi:2024cjd,Chen:2025qub,Gaberdiel:2026sfg,Kikuchi:2026kox,Ambrosino:2026umb,Benedetti:2026drn}. There exists much progress in the understanding of the unbroken subalgebraic structures, but the role of the connected etale algebra (or integer spin nonsimple currents) has rarely been discussed.

The condition to obtain the connected etale algebra is nothing but the welldefinedness of the wavefunction constructed from $\mathbf{A}_{\text{ub}} \boxtimes\overline{\mathbf{A'}}_{\text{ub}}$, and can be interpreted as the Lieb-Schultz-Mattis type anomaly free condition\cite{Lieb:1961fr,Schultz:1964fv,Cho:2017fgz}: The conformal \emph{spin} $s$ of the condensed objects $\alpha_{\text{ub}}\boxtimes \overline{\rho(\alpha_{\text{ub}})}$ satisfy relation $s_{\alpha_{\text{ub}}\boxtimes \overline{\rho(\alpha_{\text{ub}})}}\in \mathbb{Z}/2$ where $\alpha_{\text{ub}}$ is a simple object in $\mathbf{A}_{\text{ub}}$, $\mathbb{Z}$ is the set of integers. Depending on the chirality of the theories, the expression of conformal spin $s$ by conformal dimensions $h$ will be changed, such as $s_{\alpha_{\text{ub}}\boxtimes \overline{\rho(\alpha_{\text{ub}})}}=\pm h_{\alpha_{\text{ub}}}\pm h_{\rho(\alpha_{\text{ub}})}$. This (half) integer condition, $s_{\alpha_{\text{ub}}\boxtimes \overline{\rho(\alpha_{\text{ub}})}}\in \mathbb{Z}/2$, itself appears commonly in the studies on TQFTs, for example in \cite{DavydovMugerNikshychOstrik+2013+135+177,Kong:2019cuu,Kaidi:2021gbs}. We also note several works studying the corresponding structure in CFTs or SymTFTs\cite{Furuya:2015coa,Lecheminant:2015iga,Cho:2017fgz,Numasawa:2017crf,Fukusumi_2022_c,Kikuchi:2022ipr,Kaidi:2023maf,Nakayama:2024msv,Fukusumi:2025ljx,Delmastro:2025ksn}. More historically, the corresponding objects for unbroken group symmetry has been studied in the name integer spin simple current or discrete torsion\cite{Vafa:1986wx,Hamidi:1986vh,Schellekens:1990ys,Gato-Rivera:1990lxi,Gato-Rivera:1991bqv,Kreuzer:1993tf}. Whereas the (half) integral condition is evident from the values of conformal spin in the above discussions, the studies on their categorical formulations have been still in development. For example, the resultant fusion rules and modular covariants after the condensation by taking the Deligne product $\boxtimes$ have not been studied systematically when coupling two different models. We cite a pioneering work \cite{DavydovMugerNikshychOstrik+2013+135+177} introducing connected etale (or \'{e}tale) algebra in this research direction, and highlight some related works\cite{Kong:2019cuu,Zhang:2024bye,Fukusumi:2024ejk,Fukusumi:2025ljx} in condensed matter physics for further studies.

Similar to the anomaly-free condition of an unbroken subalgebra, there will exist an anomaly-free condition of preserved idempotents or sectors. We note that whereas the connected etale algebra $\mathbf{A}_{\text{ub}}$ plays a significant role in constructing coupled models, it is insufficient to determine the homomorphism uniquely\cite{Fukusumi:2025clr}. Interestingly, if one assumes the anomaly-free condition for every preserved idempotent $\mathbb{S}_{\rho,\times}$ (not anyon $\mathbf{A}_{\text{ub}}$ and $\mathbf{A'}_{\text{ub}}$) to the flow from the tricritical Ising model, $M(4,5)$, to the Ising model, $M(3,4)$, the homomorphism is unique\cite{Fukusumi:2025xrj}. Without this additional condition, there exist $4$ homomorphisms even when assuming the structure of connected etale algebra, the unbroken Ising fusion ring in this case. The fusion rule of the tricritical Ising model can be expressed as tensor product of the Ising and Fibonnacci fusion rules $M(4,5)=\text{Ising}\otimes \text{Fibonacci}=\{ I^{[1]}, \psi^{[1]}, \sigma^{[1]}\}\otimes \{ I^{[2]},\tau^{[2]}\}$. The conformal dimensions of the objects are $h_{I^{[1]}\otimes I^{[2]}}=0, h_{\psi^{[1]}\otimes I^{[2]}}=3/2, h_{\sigma^{[1]}\otimes I^{[2]}}=7/16, h_{I^{[1]}\otimes \tau^{[2]}}=3/5,h_{\psi^{[1]}\otimes \tau^{[2]}}=1/10, h_{\sigma^{[1]}\otimes \tau^{[2]}}=3/80$. On the other hand, the conformal dimensions of the Ising model, $M(3,4)=\{ I', \psi', \sigma'\}$, are $\{ 0, 1/2, 1/16\}$. Hence, the condition of the half-integral condition of the preserved idempotents is satisified in the flow $M(4,5)\rightarrow \overline{M(3,4)}$, only with $\mathbb{S}_{\rho, \times}=\{ (I^{[1]}\otimes I^{[2]}, \overline{I'}), (\psi^{[1]}\otimes I^{[2]}, \overline{\psi'}), (\sigma^{[1]}\otimes I^{[2]}, \overline{\sigma'})\}$. One can straightforwardly confirm the condition by calculating $h_{\alpha}\pm h_{\alpha'}=\mathbb{Z}/2$ for $(\alpha,\alpha')\in\mathbb{S}_{\rho,\times}$. This flow inevitably changes the chirality of anyons as in the Alice ring\cite{SCHWARZ1982141,SCHWARZ1982427}, and it is surprising that even the simplest flow shows such an exotic structure. More generally, if the condition $\mathbf{A}_{\text{ub}}=\mathbf{A'}$ is satisfied, the same determination of the homomorphism or $\mathbb{S}_{\rho,\times}$ can apply to the flow\cite{Fukusumi:2026hdh}.

\subsection{Nambu-Goldstone type phenomenology and correspondence: Operator counting of ideal, module and unbroken symmetry}
\label{Nambu-Goldstone}
It is worth reminding that the sum of the number of symmetry operator in $\mathbf{A'}$, $n'$, and that of modules of the dual massive RG, $\{ |\alpha\rangle \rangle\}_{\alpha\in\mathbb{S}_{\rho}^{\text{c}}}$ with $n-n'$ objects, are equal to that of UV symmetry operators, $n=(n-n')+n'$. It is analogous to the Nambu-Goldstone (NG) theorem\cite{Nambu:1960tm,Goldstone:1961eq,Goldstone:1962es}, and we note the corresponding phenomenology at the present stage. Historically, the connection between the pseudo-NG modes and Higgs modes has been studied in the high-energy physics phenomenology\cite{Kaplan:1983fs,Kaplan:1983sm,Georgi:1984af,Dugan:1984hq}, and we cite reviews\cite{Panico:2015jxa,Brauner:2024juy}.  We also note a review \cite{Watanabe:2019xul} for readers in condensed matter. When assuming the perturbative expression of the massless flow, the ideal $\mathbb{S}_{\rho}^{\text{c}}$ which corresponds to the massive part in the massless flow, can be interpreted as pseudo-Nambu-Goldstone modes, and they are gapped out gradually by adding the strength of the coupling constant $(\lambda>0)$ of the relevant perturbations. The slow transition property of the massless flow can be found in numerical simulations, for example, in \cite{Zou_2018}. We note the algebraic dimension of the ideal or pseudo-NG modes as $n_{\text{pNG}}=n-n'$. 

In the dual massive RG $(\lambda<0)$, which can be obtained by changing the sign of the coupling constant of perturbations deriving the massless flow, the ideal appears as a module or degenerate ground states with spontaneous symmetry breaking. In this situation, the unbroken symmetry $\mathbf{A}_{\text{ub}}$ relates these ground states to each other in the massive RG, and $\mathbf{A}_{\text{ub}}$ corresponds to the NG modes (which should be distinguished from the \emph{pseudo}-NG modes in the massless RG). In general, the algebraic dimension of $\mathbf{A}_{\text{ub}}=\mathbf{A'}_{\text{ub}}$, $\text{dim}(\mathbf{A}_{\text{ub}})$, or operator couting of the NG modes is smaller than $n'$ because of the inclusion $\mathbf{A'}_{\text{ub}}\subset \mathbf{A'}$. Hence, the following inequality holds,
\begin{equation}
n=n_{\text{pNG}}+n'\ge n_{\text{pNG}}+\text{dim}(\mathbf{A}_{\text{ub}})
\end{equation}
Because the detection of the unbroken symmetry and its action on the module are fundamental, the arguments by Graham-Watts \cite{Graham:2003nc} are important for further studies, and we note related recent papers again\cite{Thorngren:2019iar,Kikuchi:2022gfi,Kikuchi:2023gpj,Fukusumi:2024ejk,Fukusumi:2025clr,Wen:2025xka,Choi:2025ebk,Fukusumi:2025xrj,Nakayama:2024msv,Kikuchi:2024cjd,Chen:2025qub,Gaberdiel:2026sfg,Kikuchi:2026kox,Ambrosino:2026umb,Benedetti:2026drn}.

\section{Conclusion and discussion}
\label{conclusion}

In this paper, we have established a classification of anyons at the level of the fusion ring, under the existence of the Verlinde formula\cite{Verlinde:1988sn}. A general formula, Eq.\eqref{Verlinde-like_formula}, determining the homomorphism has been obtained. By applying the ideas in the generalized symmetry and related fields, we have also demonstrated requirements for the possible categorical descriptions compatible with the NG-type phenomena. Moreover, at the level of SymTFTs, we have demonstrated our method applies to fractional supersymmetric or nonlocal models based on the discussions in \cite{Fukusumi:2024cnl,Fukusumi:2024ejk,Fukusumi:2025ljx,Fukusumi:2025xrj,Fukusumi:2025fir} and obtained Eq.\eqref{sequence} and Eq.\eqref{RG_similarity}.  

For a future research direction, whereas our method provides a systematic description of the domain wall at the level of SymTFTs, their analytical properties are largely open. Even when restricting our attention to the simplest models, the corresponding phenomena are unexplored in general, as can be seen in the extended tensored models in \cite{Crnkovic:1989ug,Gaiotto:2012np,Fukusumi:2024ejk,Antinucci:2025uvj,Fukusumi:2025ljx,Furuta:2025ahl}. Related to this research direction, further studies on the conformal interface\cite{Wong:1994np,PhysRevLett.77.2604,Oshikawa:2005fh,Quella:2006de,Kimura:2014hva,Kimura:2015nka,Lopes_2020} are important. By applying the folding trick\cite{Wong:1994np} to the domain wall problems, related symmetry breaking branes\cite{Affleck:1998nq,Fuchs:1999zi,Fuchs:1999xn,Birke:1999ik,Quella:2002ct,Ishikawa:2005ea} are also important. We also note that the resultant extended models break the conventional operator-state correspondence\cite{Fukusumi:2023psx}, whereas the algebraic structure of the graded SymTFT still survives. This phenomenon is a variant of the obstruction realizing a chiral Majorana fermion\cite{Nielsen:1981hk,Nielsen:1981xu}. For this research direction, the generalization of the correspondence between nonchiral fusion ring and conformal bootsrap techique\cite{Rida:1999ru,Rida:1999xu,Fuchs:2002cm,Nivesvivat:2025odb} (or Moore-Seiberg data\cite{Moore:1988qv,Moore:1988ss,Moore:1989vd}) will become significant. Finally, we note recent attempts on the higher-dimensional models\cite{Johnson-Freyd:2020usu,Kong:2024ykr,Antinucci:2025fjp} where our method will be applicable. 

\emph{Note added}: 
After completing the present work, the author has noticed another reasonable framework for domain walls, called pro-functor. The mapping can be nonsurjective, because it requires a kind of tensor decomposition of the UV theory coming from coset or level-rank duality structures. However, in principle, the techniques in the present manuscript apply to this setting by introducing the modified version of idempotents. The connection between the massless RG and the monoidal functor is rather indirect, but the connection between the pro-functor and massless RG seems more reasonable. (See the corresponding discussion in \cite{fukusumi2026organizingtransitionscascadesgeneralized}.)

\section{Acknowledgement}

The author thanks Gerard Watts and Ingo Runkel for helpful discussions and for notifying the author of the elementary but fundamental method of solving the ring homomorphism by studying the representation by idempotents.  
The author thanks Yuma Furuta, Taishi Kawamoto, and Shinichiro Yahagi for related collaborations, and Shuma Nakashiba for the related discussions. The author thanks the support from NCTS.

\clearpage

\appendix

\section{Bulk semionization for Majorana fermion and related homomorphism: Structures compatible with homomorphism}

For the readers unfamiliar with sandwich construction, we note the consequence of the bulk semionization. For more technical and phenomenological details, see related works by the author\cite{Fukusumi:2024cnl,Fukusumi:2024ejk,Fukusumi:2025ljx}.

First the Ising bulk CFT has three objects, $\{ I, \epsilon, \sigma_{\text{Bulk}} \}$ where $I$ is the identity operator, $\epsilon$ is the energy operator and $\sigma_{\text{Bulk}}$ is the $\mathbb{Z}_{2}$ order operator. They satisfy the following fusion rule,
\begin{align}
\epsilon \times \epsilon &=I, \\
\epsilon \times \sigma_{\text{Bulk}} &=\sigma_{\text{Bulk}}, \\
\sigma_{\text{Bulk}} \times \sigma_{\text{Bulk}} &=I+\epsilon, 
\end{align}

The $\mathbb{Z}_{2}$ extension is the simple procedure, just extending the bulk Ising fusion rule to $\mathbb{Z}_{2}\otimes \mathbf{B}$. We denote the $\mathbb{Z}_{2}$ as $\{ 1, \psi \}$. Hence, the extended theory $\mathbf{F}=\mathbb{Z}_{2}\otimes \mathbf{B}$ has six objects. Bulk semionization is just a simple procedure by acting $\mathbb{Z}_{2}$ orbifolding procedure to the bulk part $\mathbf{B}$ under preserving the $\mathbb{Z}_{2}$ structure. The consequence can be written as follows,
\begin{align}
I'&=1\otimes \left(\frac{I+\epsilon}{2}\right)\\
\psi'&= \psi \otimes \left(\frac{I+\epsilon}{2}\right),\\
e' &= 1 \otimes \frac{\sigma_{\text{Bulk}}}{\sqrt{2}} ,\\
m' &=\psi \otimes \frac{\sigma_{\text{Bulk}}}{\sqrt{2}} 
\end{align}
One can check that the above four objects, $\{ I',\psi',e',m'\}$, satisfy the double semion fusion rules, and they provide the $\mathbb{Z}_{2}$-graded SymTFT $\mathbf{S}$ or chiral Majorana fusion ring\cite{Ginsparg:1988ui}. It should be stressed that $\otimes$ is \emph{not} the Deligne product, $\boxtimes$, but the tensor product. Because the above sandwich construction is just a combination of the procedures taking tensor product and subalgebra, it is compatible with the homomorphism $\rho$ in the main text when the $\mathbb{Z}_{N}$ group ring structure is unbroken. The discussions in this section hold true for the $\mathbb{Z}_{N}$ symmetric model $\mathbf{B}$, $\mathbb{Z}_{N}$ extended theory $\mathbf{F}$, and the resultant $\mathbb{Z}_{N}$-garded SymTFT $\mathbf{S}$ in general.

We also note that there exist related but different connections between two copies of the chiral Ising model and chiral Majorana fermion model\cite{Hung:2013qpa,Lan:2014uaa}. We also note that the related lattice gauge theoretic formulations have been studied in \cite{Zhao:2025zsb}. We distinguish the two copies of the chiral Ising model by upper index $\{1\}$ and $\{2\}$ and denote the UV theories as $\{ I^{i},\psi^{\{ i\}}, \sigma^{\{ i\}}\}$. Then the following ring homomorphism holds,
\begin{align} 
\rho(I^{\{1\}}\otimes I^{\{2\}})&=I', \\
\rho(\psi^{\{1\}}\otimes I^{\{2\}})&=\psi', \\
\rho(I^{\{1\}}\otimes \psi^{\{2\}})&=\psi', \\
\rho(\psi^{\{1\}}\otimes \psi^{\{2\}})&=I', \\
\rho(\sigma^{\{1\}}\otimes \sigma^{\{2\}})&=e'+m'. 
\end{align}
It should be stressed that we have taken subalegebra of the two copies of the Ising theory, excluding $\sigma^{\{1\}}\otimes I^{\{2\}}$, $I^{\{1\}}\otimes \sigma^{\{2\}}$,$\sigma^{\{1\}}\otimes \psi^{\{2\}}$, $\psi^{\{1\}}\otimes \sigma^{\{2\}}$, and this exclusion is also a consequence of half-integral condition in forming the connected etale algebra. Because the IR theory does not have the algebraic dimension of the double semion model, $4$, the homomorphism is \emph{not} surjective and unconventional from the context of massless RG and gapped domain wall. However, by introducing the relation $\sigma'=\frac{e'+m'}{\sqrt{2}}$, the homomorphism, $\rho:\otimes_{i}\mathbf{A}^{\{ i\}}\rightarrow \{I', \psi', \sigma' \}$ becomes surjective. This identification inevitably induces the unusual coefficient $\sqrt{2}$ \footnote{Moreover, this number corresponds to the quantum dimension of $\sigma'$. We study related phenomena elsewhere.}. In other words, the object $\mu=\frac{e'-m'}{\sqrt{2}}$ is \emph{emergent} from this view. However, this emergence stems from the discrepancy between the bosonic UV and fermionic IR theories, and one can exclude this emergence by treating either the bosonic UV and IR theories or the fermionic UV and IR theories.

It should be remarked that to make the homomorphism surjective in the fermionic basis, it is necessary to introduce the extended object, chiral disorder field $\mu^{\{i\}}$, which satisfies the following relations,
\begin{equation}
\rho(\mu^{\{1\}}\otimes \mu^{\{2\}})=e'-m' 
\end{equation}
where one can identify $\sigma$ and $\mu$ as $\sigma=\frac{e+m}{\sqrt{2}}$ and $\mu=\frac{e-m}{\sqrt{2}}$. This relationship between the chiral Majorana fermion and (extended) chiral Ising model has already appeared in the literature on fermionic models\cite{Ginsparg:1988ui}.

\section{Extension, orbifolding and isomorphisms}
\label{extension}
In this section, we observe the implications of the results for more general systems, beginning the discussion from the established local bosonic models. The results are summarized in Eq.\eqref{sequence} and Eq. \eqref{RG_similarity}. First, by applying the Moore-Seiberg data\cite{Moore:1988qv,Moore:1988ss,Moore:1989vd} or Bais-Slingerland anyon condensation\cite{Bais:2008ni}, the following ring isomorphism between a chiral (or antichiral) fusion ring, $\mathbf{A}$ and $\mathbf{A'}$, and a nonchiral fusion ring\cite{Fuchs:1993et,Rida:1999xu,Rida:1999ru,Nivesvivat:2025odb}, $\mathbf{B}$ and $\mathbf{B'}$, should hold,
\begin{equation}
\mathbf{B}=\mathbf{A}, \mathbf{B'}=\mathbf{A'}.
\label{isomorphisms}
\end{equation}
For the successive discussion, we denote the objects in the bulk theory as $\Phi_{\alpha}\in \mathbf{B}$ and assume the same relations for the IR theory. The above ring isomorphism can also be interpreted as a consequence of the reduction from the tensor product $\mathbf{A}\otimes \overline{\mathbf{A}}$ to the Deligne product  $\mathbf{A}\boxtimes \overline{\mathbf{A}}=\mathbf{B}$. In category theory, $\mathbf{A}$ is called a modular tensor category (MTC) and $\mathbf{B}$ is called a spherical fusion fusion category (SFC). The SFC (or Bauer-Picard groupoid\cite{Etingof:2009yvg}) is less familiar in the fields, but it provides the data of conformal bootstrap\cite{Polyakov:1974gs,Fuchs:1993et,Rida:1999xu,Rida:1999ru,Fuchs:2002cm,Nivesvivat:2025odb}. One can see well-summarized related discussions in the article of \cite{vanVliet:2025swv} by Ingo Runkel.

Hence, the homomorphism $\rho: \mathbf{A} \rightarrow \mathbf{A'}$ naturally induces the ring homomorphism between nonchiral fusion rings $\rho :\mathbf{B}\rightarrow \mathbf{B'}$ with Eq.\eqref{isomorphisms}. Next, we assume that the homomorphism does not break the $\mathbb{Z}_{N}$ group structure. In this setting, one can apply the $\mathbb{Z}_{N}$ extension of the UV and IR theory in a way compatible with the application of the homomorphism by assuming the $\mathbb{Z}_{N}$ symmetry is unbroken. First, we assume the $\mathbb{Z}_{N}$ simple current $j\in \mathbf{A}$ and $j'\in \mathbf{A'}$ are preserved under the application of $\rho$, i.e. $\rho(j)=j'$. The extended theory $\mathbf{F}$ or $\mathbf{F'}$ can be identified as $\mathbb{Z}_{N}\otimes\mathbf{B}$ or $\mathbb{Z'}_{N}\otimes\mathbf{B'}$. Hence, one can obtain the ring homomorphism of the $\mathbb{Z}_{N}$ extended nonchiral fusion ring as,
\begin{equation} 
\rho: \mathbf{F}\rightarrow \mathbf{F'}
\end{equation}
by defining the action of $\rho: \mathbb{Z}_{N}\otimes\mathbf{B}\rightarrow \mathbb{Z'}_{N}\otimes\mathbf{B'}$ as $\rho(j\otimes \Phi_{\alpha})=j'\otimes \rho(\Phi_{\alpha})$, where $\Phi_{\alpha}\in \mathbf{B}$. 

By applying the bulk semionization\cite{Fukusumi:2024cnl,Fukusumi:2024ejk,Fukusumi:2025ljx} or sandwich construction, one can obtain the corresponding homomorphism between $\mathbb{Z}_{N}$-graded SymTFTs,
\begin{equation} 
\rho: \mathbf{S}\rightarrow \mathbf{S'}
\end{equation}
by taking the subalgebra of the extended bulk theories, such as $\mathbf{S}=\mathbb{Z}_{N}\otimes (\mathbf{B}/\mathbb{Z}_{N})\subset \mathbf{F}$. For the readers interested in the technical details of this procedure, bulk semionization, we note the works by the author\cite{Fukusumi:2024cnl,Fukusumi:2024ejk,Fukusumi:2025ljx}. Technically, this operation can be performed by taking a subalgebra, $(\mathbf{B}/\mathbb{Z}_{N})\subset\mathbf{B}$, and this operation is compatible with the application of $\rho$. Consequently, one can obtain the following nontrivial sequence for the $\mathbb{Z}_{N}$ symmetric models,
\begin{equation}
\{\rho: \mathbf{A}\rightarrow \mathbf{A'} \}
\Rightarrow\{\rho: \mathbf{F}\rightarrow \mathbf{F'} \}
\Rightarrow\{\rho: \mathbf{S}\rightarrow \mathbf{S'} \}.
\label{sequence}
\end{equation}
In condensed matter terminology, the extended theories $\mathbf{F}$ or $\mathbf{S}$ correspond to the symmetry enriched systems. By studying subgroup structure of $\mathbb{Z}_{N}$, one can apply orbifolding procedures to the sequence.
One can see more intuitive or categorical arguments in the recent works\cite{Huang:2023pyk,Huang:2024ror,Bhardwaj:2024ydc,KNBalasubramanian:2025vum,Heinrich:2025wkx,Chen:2025qub,heinrich2025generalisedorbifoldsgequivariantisation} and earlier works \cite{Turaev:2000ug,mueger2004galoisextensionsbraidedtensor,Etingof:2009yvg,etingof2009weaklygrouptheoreticalsolvablefusion,Barkeshli:2014cna}, although the calculation of the fusion coefficients is sometimes difficult to read. 

We also note that by exchanging the symbol $I$ and $I'$ to some other symbol $j$ and $j'$, one can obtain the corresponding homomorphism for nonlocal models when $j$ and $j'$ are invertible satisfying the conditions $\{ S_{j,\alpha}\neq 0 \}_{\alpha}, \{ S_{j',\alpha'}\neq 0 \}_{\alpha'}$. In CFT, this operation is called the Galois shuffle\cite{Gannon:2003de}, and it corresponds to the similarity transformation in non-Hermitian physics\cite{Mostafazadeh:2001jk,Mostafazadeh:2001nr,Mostafazadeh:2002id}. Because the discussions only require some replacement of the symbols as has been established in \cite{Fukusumi:2025fir}, we note only the consequent forms of the homomorphism,
\begin{equation}
\rho(\alpha)=\sum_{\alpha', (\beta_{\text{pr}}, \beta'_{\text{pr}})\in \mathbb{S}_{\rho,\times}}\frac{S_{\alpha,\beta_{\text{pr}}}S_{j',\beta'_{\text{pr}}}\overline{S_{\beta'_{\text{pr}},\alpha'}} }{S_{j, \beta_{\text{pr}}}}\alpha'.
\label{RG_similarity}
\end{equation}
One can also combine the above formula with extension or orbifolding, and study the hierarchical structure of the nonlocal fractional supersymmetric models and their orbifolded (or partially bosonized) models.

\bibliography{algebra_sym}

\end{document}